\documentclass[aps,prd,groupedaddress,showpacs,superscriptaddress,twocolumn,eqsecnum,floatfix]{revtex4-1}
\usepackage{graphicx}
\usepackage{dcolumn}
\usepackage{bm}
\begin{document}

\title {$\phi\phi$ and $J/\psi\phi$ mass spectra in decay $B^0_s\to J/\psi\phi\phi$.}

\author{A.~A.~Kozhevnikov}
\email[]{kozhev@math.nsc.ru} \affiliation{Laboratory of
Theoretical Physics, S.~L.~Sobolev Institute for Mathematics, Novosibirsk, Russian
Federation}
\affiliation{Novosibirsk State University, Novosibirsk, Russian
Federation}

\date{\today}

\begin{abstract}
The mass spectra of the $\phi\phi$ and $J/\psi\phi$  states in the
decay $B^0_s\to J/\psi\phi\phi$ recently observed by LHCb are
calculated  in the model which takes into account the
$J^P=0^+,0^-,2^+$ intermediate resonances $R_1$, $R_2$ in the
$\phi\phi$ channel and the $J^P=1^+$ ones, $X_1$, $X_2$, in the
$J/\psi\phi$ channel.  When obtaining the expressions for the
effective amplitudes and mass spectra,  the approximate threshold
kinematics of the decay is used essentially. The $R_1-R_2$ and
$X_1-X_2$ mixings arising due to the common decay modes $\phi\phi$
and $J/\psi\phi$, respectively, are also taken into account. The
obtained expressions for the mass spectra are applied for
extracting  the information about masses and coupling constants of
the resonances in the $\phi\phi$ and $J/\psi\phi$ final states.
\end{abstract}
\pacs{13.25.Hw,14.40.Cs,14.40.Nd}

\maketitle

\section{Introduction}
\label{intro}

Recently, the LHCb collaboration has reported the observation of
the decay $B^0_s\to J/\psi\phi\phi$ \cite{LHCb}. The interest in
this decay is related  to the possible  existence of the exotic
glueball state  decaying into the $\phi\phi$ pair
\cite{Linden,Etkin85,Booth86,Etkin88}. The spin-parity quantum
numbers of the resonance states decaying into $\phi\phi$ are
reported to be $J^P=0^+ , 0^-$, and $2^+$ \cite{pdg,BESIII}.

The LHCb collaboration has also reported the observation of four
resonance structures in the decay $B^+\to J/\psi\phi K^+$,   in
the mass range 4140 -- 4700 MeV,   decaying into $J/\psi\phi$
\cite{LHCb1,LHCb2}. This group of resonances is  widely discussed
because of their possible exotic nature \cite{aa16,zhu16,mai16},
side by side with the explanations based on the dynamical
rescattering effects \cite{liu16}. Two of them, $X(4140)$ and
$X(4274)$, have masses in the range 4100 -- 4350 MeV attained in the
$J/\psi\phi$ mass spectrum in the decay $B^0_s\to J/\psi\phi\phi$
\cite{LHCb}. The preferable spin-parity quantum numbers of these
resonances are $J^P=1^+$ \cite{LHCb1}.

The data of Ref.~\cite{LHCb} were plotted against the phase space
distribution which  was shown to be inadequate because the
resonance bumps were seen in both the $\phi\phi$ and $J/\psi\phi$
mass spectra \cite{LHCb}. The aim of the present work is to write
down the amplitudes and the mass spectra of the above final states
in the decay $B^0_s\to J/\psi\phi\phi$ upon taking into account
possible intermediate resonance states in the $\phi\phi$ and
$J/\psi\phi$ channels, irrespective of the model assumptions about
their nature.

The task of obtaining the effective amplitudes is greatly simplified when one takes into account the near-threshold kinematics of the decay $B^0_s\to J/\psi\phi\phi$. Then one can neglect the higher partial waves in the decay amplitudes. The expressions for the $\phi\phi$ and $J/\psi\phi$ mass spectra obtained under such an approximation upon taking into account the intermediate resonances $R_{1,2}\to\phi\phi$ and $X_{1,2}\to J/\psi\phi$ are used to extract from the fits the masses and coupling constants  of these resonances. It should be emphasized that the presentation of the results in terms of coupling constants is more informative than the popular representation in terms of the partial widths.

The paper is organized as follows. Section \ref{sec2} contains the expressions for the lowest momenta effective vertices of the $B^0_s\to R_{1,2}J/\psi$,  $B^0_s\to X_{1,2}\phi$, $R_{1,2}\to\phi\phi$, and $X_{1,2}\to J/\psi\phi$ transitions assuming $J^P=0^+,0^-,2^+$ for the $R_{1,2}$ resonances and $J^P=1^+$ for the $X_{1,2}$ ones. The partial decay widths $R_{1,2}\to\phi\phi$ and $X_{1,2}\to j/\psi$ are given there. Section~\ref{amps} is devoted to the derivation of the $B^0_s\to J/\psi\phi\phi$ decay amplitudes upon taking into account the $R_1-R_2$ mixing due to the common $\phi\phi$ decay channel, and $X_1-X_2$ mixing due to the $J/\psi\phi$ one. The modulus squared of the $B^0_s\to J/\psi\phi\phi$ decay amplitude and the expressions for the $\phi\phi$ and $J/\psi\phi$ mass spectra are calculated in Sec.~\ref{spectra}. In Sec.~\ref{sec5}, these expression are applied to the description of the LHCb data \cite{LHCb}. Section~\ref{sec6} contains the brief discussion of the obtained results. Section~\ref{sec7} serves  as a conclusion. Some details used in the derivation of expressions in the main text are given in the Appendices.

\section{Effective vertices and partial decay widths}
\label{sec2}
~

We assume the existence of the resonances at $m_{\phi\phi}=2.07$
and 2.2 GeV in the $\phi\phi$ mass spectrum which will be called
as $R_{1,2}$, and the resonances  $X(4140)$ and $X(4274)$
with  masses in the range 4100 -- 4350 MeV, in the $J/\psi\phi$ mass
spectrum \cite{CDF,LHCb12} which will be called $X_{1,2}$. One
needs the effective vertices $B^0_s\to J/\psi R$, $R\to\phi\phi$,
$B^0_s\to\phi X$,  $X\to J/\psi\phi$. See Fig.~\ref{fig1}. Since
all particles in the final state of the decay $B^0_s\to
J/\psi\phi\phi$ have unit spin, the number of effective
contributions to the decay amplitude is frighteningly large,
especially when taking into account the fact that space parity is
not conserved because the $B^0_s$ meson decays  due to the weak
interactions, so that only the angular momentum conservation, not
the space parity, restricts the number of possible Lorenz
structures in the effective decay amplitudes. The couplings of $R$
with $\phi\phi$ and $X$ with $J/\psi\phi$ are considered to be due
to the strong interactions, hence they conserve parity, but again
the unit spin admits many independent effective contributions. The
situation can be greatly simplified if one takes into account the
fact that the kinematics of the decay
\begin{equation}\label{dec}
  B^0_s(Q)\to J/\psi(q)+\phi(k_1)+\phi(k_2)\end{equation}
 is such that all particles in the final state have relatively low momenta. Indeed, the invariant mass of the $\phi\phi$ pair varies in the range
 \begin{equation}\label{m12}
   2m_\phi\leq m_{\phi\phi}\leq m_{B^0_s}-m_{J/\psi},
 \end{equation}
 that is, $2.04\leq m_{\phi\phi}\leq 2.27$ GeV. The maximum momentum of the $\phi$ is reached when the final $J/\psi$ meson is at rest while two $\phi$ mesons move in opposite directions, and this gives $|{\bm k}_1|_{\rm max}/m_\phi=0.5$. Analogously, the invariant mass of the $J/\psi\phi$ state varies in the range
 \begin{equation}\label{m13}
   m_\phi+m_{J/\psi}\leq m_{J/\psi\phi}\leq m_{B^0_s}-m_\phi,
 \end{equation} that is, $4.12\leq m_{J/\psi\phi}\leq 4.35$ GeV. The maximum momentum of the $J/\psi$ meson is reached when  one of the $\phi$ mesons is at rest, while other moves oppositely to $J/\psi$ resulting in $|{\bm q}|_{\rm max}/m_{J/\psi}=0.2$.  However, these relatively small ratios are not in fact reached because the above kinematic situations taking place at the border of phase space are suppressed by the final state phase space factors. See Eqs.~(\ref{s12}) and (\ref{s23}) below. This permits one to take into account only the effective decay vertices with the lowest nonvanishing powers of momenta.

Let us start with the parity-conserving effective vertices of $R\to\phi\phi$ and $X\to J/\psi\phi$.
The required  lowest momenta $R\to\phi\phi$ vertices for various quantum numbers are
\begin{eqnarray}\label{Rvert}
M_{R(0^+)\to\phi\phi}&=&g_{R\phi\phi}(\epsilon^{(\phi_1)}\cdot\epsilon^{(\phi_2)}),\nonumber\\
M_{R(0^-)\to\phi\phi}&=&g_{R\phi\phi}\varepsilon_{\mu\nu\lambda\sigma}k_{1\mu}\epsilon^{(\phi_1)}_\nu k_{2\lambda}\epsilon^{(\phi_2)}_\sigma,\nonumber\\
M_{R(2^+)\to\phi\phi}&=&g_{R\phi\phi}T_{\mu\nu}\epsilon^{(\phi_1)}_\mu\epsilon^{(\phi_2)}_\nu,
\end{eqnarray}
where $T_{\mu\nu}$ is the polarization tensor of the spin two resonance, $\epsilon^{(\phi_{1,2})}$ and $k_{1,2}$ stand for the polarization four-vector and four-momentum of the $\phi_{1,2}$ meson, and $\varepsilon_{\mu\nu\lambda\sigma}$ is the totally  antisymmetric unit Levy-Civit\`{a} tensor.

The quantum numbers of the $X(4140)$ and $X(4274)$ resonances are
now established: $J^{PC}=1^{++}$ \cite{LHCb1,LHCb2}.  Then the
effective lowest momentum $X\to J/\psi\phi$ vertex looks like
\begin{eqnarray}\label{Xvert}
M_{X(1^+)\to
J/\psi\phi}&=&g_{XJ\psi\phi}\varepsilon_{\mu\nu\lambda\sigma}p_\mu\epsilon^{(X)}_\nu\epsilon^{(J/\psi)}_\lambda\epsilon_\sigma^{(\phi)}.
\end{eqnarray}In the above expressions, $\epsilon^{(X)}$, $\epsilon^{(J/\psi)}$  stand for the polarization four-vectors of the $X$, $J/\psi$ mesons, respectively,  and $p_\mu$ is the four-momentum of the $X$ resonance. The justification of this expression is given in the Appendix \ref{B}. All effective vertices are assumed to be real, and the possible dependence on the momentum squared is neglected.

Let us give the effective amplitudes  for the weak decays of the $B^0_s$ meson. The lowest momenta amplitudes of the transitions $B^0_s\to J/\psi R$ for different quantum numbers of the resonance $R$ are
\begin{eqnarray}\label{BR}
M_{B^0_s\to J/\psi R(0^\pm)}&=&2g_{B^0_sJ/\psi R}(\epsilon^{(J/\psi)}\cdot k),\nonumber\\
M_{B^0_s\to J/\psi R(2^+)}&=&g_{B^0_sJ/\psi
R}T_{\mu\alpha}\epsilon^{(J/\psi)}_\mu q_\alpha,
\end{eqnarray}where $k=k_1+k_2$. Note that, in the first expression in Eq.~(\ref{BR}), the amplitude conserves (breaks) the space parity in the case of
$R$ with $J^P=0^-$ ($J^P=0^+$), respectively. The expression of the decay amplitude  to the tensor resonance in the second line of Eq.~(\ref{BR}) breaks parity. The parity-conserving amplitude is $\propto
T_{\mu\alpha}\varepsilon_{\mu\nu\lambda\sigma}q_\alpha\epsilon^{(J/\psi)}_\nu
k_\lambda q_\sigma$, hence it has the $D$ wave form to be neglected side by side with the parity-breaking $D$ wave expression $\propto T_{\mu\alpha} q_\mu
q_\alpha(\epsilon^{(J/\psi)}\cdot k)$. The lowest momentum $B^0_s\to\phi X$ transition amplitude,
\begin{equation}\label{MBX}M_{B^0_s\to\phi X(1^+)}
=g_{B^0_sX\phi}(\epsilon^{(\phi)}\cdot\epsilon^{(X)}),\end{equation}
breaks space parity.

Our goal here is to take into account the energy dependence of the partial widths of the resonances involved, as well as their mixing due to the common decay modes (if any). The Particle Data Group (PDG) gives the $\phi\phi$, $K\bar K$ decay modes for the tensor $f_2(2010)$, $f_2(2300)$, and $f_2(2340)$ resonances and the $\eta\eta$ one for the $f_2(2340)$ resonance \cite{pdg}. Again, because of the low statistics of the available data we will take
into account only the  $\phi\phi$ decay mode relevant in the context of the data presented in Ref.~\cite{LHCb}. The same assumption will be adopted for the scalar and pseudoscalar ones observed by the BESIII collaboration \cite{BESIII}.

The standard calculation gives the  partial decay widths for the resonances with the given quantum numbers. Taking into account only the lowest nonvanishing momenta, one gets, using the effective vertices Eq.~(\ref{Rvert}), the following expressions:
\begin{eqnarray}\label{Rwidth}
\Gamma_{R(0^+)\to\phi\phi}(m^2)&=&\frac{3g^2_{R(0^+)\phi\phi}}{32\pi m^3}\lambda^{1/2}(m^2,m^2_\phi,m^2_\phi)\nonumber\\
\Gamma_{R(0^-)\to\phi\phi}(m^2)&=&\frac{g^2_{R(0^-)\phi\phi}}{64\pi m^3}\lambda^{3/2}(m^2,m^2_\phi,m^2_\phi)\nonumber\\
\Gamma_{R(2^+)\to\phi\phi}(m^2)&=&\frac{g^2_{R(2^+)\phi\phi}}{32\pi m^3}\lambda^{1/2}(m^2,m^2_\phi,m^2_\phi).
\end{eqnarray}Hereafter,
\begin{equation}\label{lambda}
\lambda(x,y,z)=x^2+y^2+z^2-2xy-2xz-2yz
\end{equation}is the standard K\"{a}ll\'{e}n function.  Correspondingly, the $X(1^+)\to J/\psi\phi$ partial width calculated from Eq.~(\ref{Xvert}) is
\begin{equation}\label{Xwidth}
  \Gamma_{X\to J/\psi\phi}(m^2)=\frac{g^2_{XJ/\psi\phi}}{8\pi m}\lambda^{1/2}(m^2,m^2_{J/\psi},m^2_\phi).
\end{equation}In what follows, the $B^0_s\to J/\psi\phi\phi$ decay amplitudes to be derived below will include only the resonance contribution with specific $J^P$.
By this reason,  all  resonances $R$  with different quantum
numbers will be labeled by the single letter $R$ in the coupling
constants, without pointing to the $J^P$ quantum numbers.

\section{Amplitudes of the decay $B^0_s\to J/\psi\phi\phi$}
\label{amps}
~

\begin{figure}
\includegraphics[width=8cm]{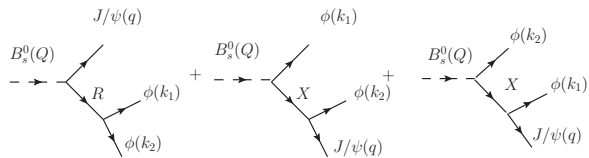}
\caption{\label{fig1}The diagrams of the decay
$B^0_s\to J/\psi\phi\phi$.}\end{figure}

According to the diagrams shown in Fig.~\ref{fig1}, let us represent the $B^0_s\to J/\psi\phi\phi$ decay amplitude as the sum of the $R$ and $X$ resonance contributions, $$M=M_R+M_X.$$
The LHCb data \cite{LHCb} visibly demonstrate the appearance of
two enhancements in the $\phi\phi$ and $J/\psi\phi$ mass spectra.
So, it will be assumed in what follows that there are two
resonances $R_1$, $R_2$  with the spin zero and (or) spin two  in
the $\phi\phi$ system mass range from 2.0 to 2.4 GeV
\cite{pdg,BESIII} and two resonances $X_1$, $X_2$ with $J^P=1^+$
in the $J/\psi\phi$ system mass range from 4.14 to 4.35 GeV
\cite{LHCb1,LHCb2}. Their masses are close, and the resonances
have common decay modes like $\phi\phi$ in the case of $R_{1,2}$
or $J/\psi\phi$ in the case of $X_{1,2}$. They can mix inside each
group. Using Refs.~\cite{ach97,ach00}  we take into account the
$R_1-R_2$ and $X_1-X_2$ mixing  by introducing the nondiagonal
polarization operators $\Pi^{(R)}_{12}$ and $\Pi^{(X)}_{12}$,
respectively. Then the simplest Breit-Wigner resonance
contribution in the case of, say, $R_{1,2}$,
\begin{equation}\label{BW}
BW\propto\frac{g_{B^0_sJ/\psi
R}g_{R\phi\phi}}{m^2_R-m^2-im_R\Gamma_R},\end{equation} should be
generalized to include the effects of energy dependent widths and
mixing by means of introducing the amplitude $G^{(R)}_{12}$ with
amputated kinematic factors (to be included below). Taking into
account the effect of resonance mixing in general case is outlined
in Appendix \ref{B} by Eq.~(\ref{Ggen}). In the case of two mixed
resonances Eq.~(\ref{Ggen}) reduces to
\begin{widetext}
\begin{eqnarray}\label{GR12}
G^{(R)}_{12}\equiv G^{(R)}_{12}(m^2)=\left(
            \begin{array}{cc}
              g_{B^0_sJ/\psi R_1}, & g_{B^0_sJ/\psi R_2} \\
            \end{array}
          \right)\left(\begin{array}{cc}
                       D_{R_2} & \Pi^{(R)}_{12} \\
                       \Pi^{(R)}_{12} & D_{R_1} \\
                                             \end{array}
                                           \right)\left(
                                                    \begin{array}{c}
                                                      g_{R_1\phi\phi} \\
                                                      g_{R_2\phi\phi} \\
                                                    \end{array}
                                                  \right)\frac{1}{D_{R_1}D_{R_2}-\Pi^{(R)2}_{12}}.
\end{eqnarray}
\end{widetext}
Here,
\begin{equation}\label{DRi}
  D_{R_i}\equiv D_{R_i}(m^2)=m^2_{R_i}-m^2-im\Gamma_{R_i\to\phi\phi}(m^2),
\end{equation}with $i=1,2$ where $\Gamma_{R_i\to\phi\phi}(m^2)$ for different quantum numbers of the $R_{1,2}$ resonances are given by Eq.~(\ref{Rwidth}), and the nondiagonal polarization operator which includes the common $\phi\phi$ mode is
\begin{eqnarray}\label{PiR12}
  \Pi^{(R)}_{12}&\equiv&\Pi^{(R)}_{12}(m^2)={\rm Re}\Pi^{(R)}_{12}(m^2)+\nonumber\\&&im\Gamma_{R_1\to\phi\phi}(m^2)\frac{g_{R_2\phi\phi}}{g_{R_1\phi\phi}}.
\end{eqnarray}
Analogously, the mixing of $X$ resonances is taken into account by introducing the amplitude with amputated kinematical factors:
\begin{widetext}
\begin{eqnarray}\label{GX12}
G^{(X)}_{12}\equiv G^{(X)}_{12}(m^2)=\left(
            \begin{array}{cc}
              g_{B^0_s\phi X_1}, & g_{B^0_s\phi X_2} \\
            \end{array}
          \right)\left(\begin{array}{cc}
                       D_{X_2} & \Pi^{(X)}_{12} \\
                       \Pi^{(X)}_{12} & D_{X_1} \\
                                             \end{array}
                                           \right)\left(
                                                    \begin{array}{c}
                                                      g_{X_1J/\psi\phi} \\
                                                      g_{X_2J/\psi\phi} \\
                                                    \end{array}
                                                  \right)\frac{1}{D_{X_1}D_{X_2}-\Pi^{(X)2}_{12}}.
\end{eqnarray}
\end{widetext}
The inverse propagator of the $X_i$ resonances ($i=1,2$) that appears in Eq.~(\ref{GX12}) is
\begin{equation}\label{DXi}
  D_{X_i}\equiv D_{X_i}(m^2)=m^2_{X_i}-m^2-im\Gamma_{X_i\to J/\psi\phi}(m^2),
\end{equation}where $m_{X_i}$ is the mass of the $X_i$ resonance. The nondiagonal polarization operator
responsible for $X_1-X_2$ mixing is written analogously to
Eq.~(\ref{PiR12}):
\begin{eqnarray}\label{PiX12}
  \Pi^{(X)}_{12}&\equiv&\Pi^{(X)}_{12}(m^2)={\rm Re}\Pi^{(X)}_{12}(m^2)+\nonumber\\&&im\Gamma_{X_1\to J/\psi\phi}(m^2)\frac{g_{X_2 J/\psi\phi}}{g_{X_1 J/\psi\phi}}.
\end{eqnarray}
The partial width of the $X_i$ decay to $J/\psi\phi$ is given by
Eq.~(\ref{Xwidth}). Equations (\ref{GR12}) and
(\ref{GX12}) reduce to the simple sum of two Breit-Wigner
contributions in the case of vanishing mixing
$\Pi^{(R,X)}_{12}\to0$. The explicit  alternative expression for
the mixing amplitude  valid in the  case of the two  resonances
mixed  via the single common decay channel is given by
Eq.~(\ref{Gpart}) in Appendix \ref{B}.

As for the Eqs.~(\ref{DRi}) and (\ref{PiR12}), their imaginary parts originated from the $\phi\phi$ loop contribution  are fixed by the unitarity relation. The real parts  are divergent when calculated from the dispersion relation upon neglecting the vertex form factors,
\begin{equation}\label{disprel}
  \Pi^{(R)}_{12}(s)=\frac{g_{R_2\phi\phi}}{\pi g_{R_1\phi\phi}}
  \int_{4m^2_\phi}^\infty\frac{\sqrt{s^\prime}\Gamma_{R_1\to\phi\phi}(s^\prime)}{s^\prime-s-i0}ds^\prime.
\end{equation}
In the case of $J^P=0^+$ and $2^+$, when restricting to the $S$-wave approximation, the divergence is logarithmic and can be regularized by the subtraction at the resonance mass. However, there is the $D$ wave contribution neglected in Eq.~(\ref{Rwidth}) which makes the divergence much stronger, and there are no model independent ways to fix the subtraction constants. The same holds for the $P$ wave decay of the $J^P=0^-$ resonance. Alternatively, one may insert the vertex form factors to make the loop integrations finite, but this again requires the fixing of additional free parameters characterizing the above form factors. The same refers to the $X$ resonances whose decay width contains the $D$-wave contribution essential in the dispersion integral at large momenta.

In practice, one can adopt the following way of action. In the present case, there are no sharp energy dependencies of the loop effects like those observed in the case of the difference of the $K^+K^-$ and $K^0\bar K^0$ loop contributions \cite{ach79,ach16}. The decay kinematics is such that it involves relatively narrow intervals of the invariant masses of the $\phi\phi$ and $J/\psi\phi$ states. See Eqs.~(\ref{m12}) and (\ref{m13}). Hence, one can ignore the possible dependence on energy of the really incalculable smooth real parts of the loop contributions and take them as constants. In the case of the diagonal polarization operators, the constants are absorbed in the masses of the resonances $m_{R_{1,2}}$, while in the nondiagonal one, $\Pi^{(R,X)}_{12}$, it is included as the free parameter $a^{(R,X)}_{12}\equiv{\rm Re}\Pi^{(R,X)}_{12}$, all these to be determined from the fit.

After these remarks, one can write the contribution of the mixed resonances in the $\phi\phi$ state.

(a) $R_{1,2}=0^+$.
\begin{equation}\label{MA}
M_R=2G^{(R)}_{12}(\epsilon^{(J/\psi)}\cdot
k)(\epsilon^{(\phi_1)}\cdot\epsilon^{(\phi_2)}),
\end{equation} which is odd under the parity inversion. Hereafter
$k=k_1+k_2$.

(b) $R_{1,2}=0^-$.
\begin{equation}\label{MC}
M_R=2G^{(R)}_{12}(\epsilon^{(J/\psi)}\cdot
k)\varepsilon_{\mu\nu\lambda\sigma}k_{1\mu}\epsilon^{(\phi_1)}_\nu
k_{2\lambda}\epsilon^{(\phi_2)}_\sigma,
\end{equation}which is even under the parity inversion.

(c) $R_{1,2}=2^+$.
\begin{equation}\label{ME}
  M_R=2G^{(R)}_{12}P_{\mu\nu,\lambda\sigma}\epsilon^{(J/\psi)}_\mu q_\nu\epsilon^{(\phi_1)}_\lambda\epsilon^{(\phi_2)}_\sigma,
\end{equation}where
\begin{equation}\label{P4}
  P_{\mu\nu,\lambda\sigma}=\frac{1}{2}\left(P_{\mu\lambda}P_{\nu\sigma}+P_{\mu\sigma}P_{\nu\lambda}\right)-\frac{1}{3}P_{\mu\nu}P_{\lambda\sigma},
\end{equation}with
\begin{equation}\label{P2}
  P_{\mu\nu}\equiv P_{\mu\nu}(k)=-\eta_{\mu\nu}+\frac{k_\mu k_\nu}{k^2}
\end{equation}stands for the result of summation over the polarizations of the intermediate tensor resonance; $\eta_{\mu\nu}={\rm diag}(1,-1,-1,-1)$.
When calculating mass spectra, it is useful to do this in the rest reference frame of the  $\phi\phi$ pair, $k=(m_{12},0,0,0)$, where
\begin{equation}\label{sm12}
m^2_{12}=(k_1+k_2)^2\end{equation} is the invariant mass squared
of the $\phi\phi$ state. In this frame, $P_{\mu\nu,\lambda\sigma}$
reduces to the three-dimensional form expressed through the
Kronecker delta:
\begin{equation}\label{P43D}
  P_{mn,ls}=\frac{1}{2}\left(\delta_{ml}\delta_{ns}+\delta_{ms}\delta_{nl}\right)-\frac{1}{3}\delta_{mn}\delta_{ls}.
\end{equation}

The second necessary ingredient for obtaining the $\phi\phi$ and $J/\psi\phi$ mass spectra in the decay $B^0_s\to J/\psi\phi\phi$ is the contribution of the $X$ exchange schematically depicted as the second and third Feynman diagrams in Fig.~\ref{fig1}. It looks like
\begin{eqnarray}\label{MX}
  M_X&=&\varepsilon_{\mu\nu\lambda\sigma}\epsilon^{(\phi_1)}_\nu\epsilon^{(J/\psi)}_\lambda\epsilon^{(\phi_2)}_\sigma
  \left[G^{(X)}_{12}(m^2_{13})(q+k_1)_\mu-\right.\nonumber\\&&\left.G^{(X)}_{12}(m^2_{23})(q+k_2)_\mu\right],
\end{eqnarray}where
\begin{equation}\label{m213}
  m_{13}^2=(q+k_1)^2
\end{equation}
and
\begin{equation}\label{m223}
  m_{23}^2=(q+k_2)^2
\end{equation}
stand for  the invariant mass squared of the $J/\psi\phi_{1,2}$ states, respectively. Note that the amplitude (\ref{MX}) is even under parity reflection and symmetric under permutation of two $\phi$ mesons. Equation (\ref{MX}) can be simplified when taking into account the small momenta of final particles. It is composed as the difference of two Lorenz-invariant expressions each of which can be evaluated in its respective rest reference frame:
\begin{eqnarray}\label{MXa}
M_X&=&m_{13}G^{(X)}_{12}(m^2_{13})\left([{\bm\epsilon}_1\times{\bm\epsilon}_2]\cdot{\bm\epsilon}^{(J/\psi)}\right)_{13}-\nonumber\\&&
m_{23}G^{(X)}_{12}(m^2_{23})\left([{\bm\epsilon}_1\times{\bm\epsilon}_2]\cdot{\bm\epsilon}^{(J/\psi)}\right)_{23},
\end{eqnarray}
where indices 13 or 23 at the vector structures point to the rest reference frame of the state $J/\psi\phi_1$ or $J/\psi\phi_2$. Now, the three-dimensional polarization vector ${\bm\epsilon}$ of the vector meson with the four-momentum $(E,{\bm p})$ is expressed through its counterpart ${\bm\xi}$ in the rest frame:
\begin{equation}\label{eps}
{\bm\epsilon}={\bm\xi}+\frac{{\bm p}({\bm\xi}\cdot{\bm p})}{m(E+m)}.
\end{equation}
Hence, both three-dimensional polarization structures in Eq.~(\ref{MXa})   can be represented in the form which includes only the polarization three-vectors in the rest frame,
$$[{\bm\epsilon}_1\times{\bm\epsilon}_2]\cdot{\bm\epsilon}^{(J/\psi)}\approx
[{\bm\xi}_1\times{\bm\xi}_2]\cdot{\bm\xi}^{(J/\psi)},$$
because they  differ by the terms  squared in momenta which can be  neglected in the considered case. Under this approximation the amplitude of $X$ exchange looks like
\begin{eqnarray}\label{MXappr}
M_X&\approx&\left[G^{(X)}_{12}(m^2_{13})m_{13}-G^{(X)}_{12}(m^2_{23})m_{23}\right]\times\nonumber\\&&[{\bm\xi}_1\times{\bm\xi}_2]\cdot{\bm\xi}^{(J/\psi)}.
\end{eqnarray}
The expressions for the $\phi\phi$ and $J/\psi\phi$ mass spectra
in the decay $B^0_s\to J/\psi\phi\phi$ are given in the next
section.

\section{Amplitudes squared and  mass spectra in the decay $B^0_s\to J/\psi\phi\phi$}\label{spectra}
~

When calculating the modulus squared of the relevant amplitude, $|M|^2\equiv |M(m^2_{12},m^2_{13},m^2_{23})|^2$, where the invariant masses squared in Eqs.~(\ref{sm12}), (\ref{m213}), and (\ref{m223}) are subjected to the constraint
\begin{equation}\label{Sigma}
m^2_{12}+m^2_{13}+m^2_{23}=m^2_{B^0_s}+m^2_{J/\psi}+2m^2_\phi\equiv\Sigma,
\end{equation}one should take into account the approximately nonrelativistic character of the problem and keep only the lowest powers of the final particle momenta. In this case, the four-dimensional scalar product of two four-momenta is
\begin{eqnarray}\label{p12}
  p_1\cdot p_2&\approx&m_1m_2+\frac{1}{2}\left(\frac{{\bm p}_1}{m_1}-\frac{{\bm p}_2}{m_2}\right)^2\approx\nonumber\\&& m_1m_2+O\left(\frac{{\bm p}^2}{m^2_{1,2}}\right).
\end{eqnarray}Then one obtains the required expressions for the specific quantum numbers of the resonance in the $\phi\phi$ state. They are the following.

(a) $R_{1,2}=0^+$. In this case $M_R$ is odd while $M_X$ is even under the space parity reflection hence they do not interfere. The modulus squared of the decay amplitude looks like
\begin{eqnarray}\label{MA2}
|M|^2&\approx&3\left|G^{(R)}_{12}(m^2_{12})\right|^2\frac{\lambda(m^2_{B^0_s},m^2_{J/\psi},m^2_{12})}{m^2_{J/\psi}}+\nonumber\\&&|M_X|^2.
\end{eqnarray}

(b) $R_{1,2}=0^-$. Here  both $R$ and $X$ contributions are even under parity reflection, hence the interference is nonzero. The expression for the modulus squared of the decay amplitude looks like
\begin{widetext}
\begin{eqnarray}\label{MB2}
|M|^2&\approx&\frac{\left|G^{(R)}_{12}(m^2_{12})\right|^2}{2m^2_{J/\psi}}\times\lambda(m^2_{B^0_s},m^2_{J/\psi},m^2_{12})
\lambda(m^2_{12},m^2_\phi,m^2_\phi)+|M_X|^2+\nonumber\\&&\frac{2m^2_{12}}{m_{J/\psi}}(m^2_{13}-m^2_{23}){\rm
Re}\left\{G^{(R)\ast}_{12}(m^2_{12})\left[G^{(X)}_{12}(m^2_{13})m_{13}-G^{(X)}_{12}(m^2_{23})m_{23}\right]\right\}.
\end{eqnarray}
\end{widetext}
Note also that the lowest order even parity $R$ contribution results in the $\phi$ mesons in $P$ wave, and by this reason it contains additional factor $\lambda(m^2_{12},m^2_\phi,m^2_\phi)$ proportional to the momentum squared of the final $\phi$ meson.

(c) $R_{1,2}=2^+$. Similar to  case (a) above, here $R$ and $X$ contributions do not interfere because they have opposite space parity. The modulus squared of the decay amplitude is
\begin{eqnarray}\label{MC2}
|M|^2&\approx&\frac{5\left|G^{(R)}_{12}(m^2_{12})\right|^2}{3m^2_{J/\psi}}\times\lambda(m^2_{B^0_s},m^2_{J/\psi},m^2_{12})+\nonumber\\&&|M_X|^2.
\end{eqnarray}
In the above expressions,
\begin{eqnarray}\label{MX2}
|M_X|^2&\equiv&|M_X|^2(m^2_{13},m^2_{23})\approx\nonumber\\&&6\left|G^{(X)}_{12}(m^2_{13})m_{13}-G^{(X)}_{12}(m^2_{23})m_{23}\right|^2
\end{eqnarray}
stands for the contribution of the intermediate $X$ resonance with
quantum numbers $J^P=1^+$. It should be emphasized once again that
only the lowest nonvanishing powers of the particle momenta are
taken into account in Eqs.~(\ref{MA2}), (\ref{MB2}), (\ref{MC2}),
and (\ref{MX2}).

The spectra of interest in the decay $B^0_s\to J/\psi\phi\phi$ are given by the following expressions. The $\phi\phi$ spectrum is
\begin{eqnarray}\label{s12}
\frac{d\Gamma}{dm_{12}}&=&\frac{\lambda^{1/2}(m^2_{B^0_s},m^2_{J/\psi},m^2_{12})\lambda^{1/2}(m^2_{12},m^2_\phi,m^2_\phi)}
{(2\pi)^3\times64m^3_{B^0_s}m_{12}}\times\nonumber\\&&\int_{-1}^1|M(m^2_{12},m^2_{13},m^2_{23})|^2dx,
\end{eqnarray}where  the explicit expression for $m^2_{13}$ to be inserted into Eq.~(\ref{s12}), in the rest frame of the $\phi\phi$ pair, is
\begin{eqnarray}\label{m213a}
 m^2_{13}&=&\frac{1}{2}(\Sigma-m^2_{12})-\frac{x}{2m^2_{12}}
 \lambda^{1/2}(m^2_{B^0_s},m^2_{J/\psi},m^2_{12})\times\nonumber\\&&\lambda^{1/2}(m^2_{12},m^2_\phi,m^2_\phi).
\end{eqnarray}Here, $x$ is the cosine of the angle between the directions of one of the $\phi$ mesons, say $\phi_1$, and the $J/\psi$ meson, in the rest frame of the $\phi\phi$ pair. The expression for $m^2_{23}$ is obtained from Eq.~(\ref{m213a}) by inverting the sign of $x$.

The expression for the $J/\psi\phi$ mass spectrum is given by the expression
\begin{eqnarray}\label{s23}
\frac{d\Gamma}{dm_{23}}&=&\frac{\lambda^{1/2}(m^2_{B^0_s},m^2_\phi,m^2_{23})\lambda^{1/2}(m^2_{23},m^2_{J/\psi},m^2_\phi)}
{(2\pi)^3\times64m^3_{B^0_s}m_{23}}\times\nonumber\\&&\int_{-1}^1|M(m^2_{12},m^2_{13},m^2_{23})|^2dx^\ast,
\end{eqnarray}where one should insert
\begin{equation}\label{m212a}
m^2_{12}=2(m^2_\phi+E_1E_2-|{\bm k}_1||{\bm k}_2|x^\ast),
\end{equation}
with
\begin{eqnarray}\label{kinem23}
E_1&=&\frac{m^2_{B^0_s}-m^2_\phi-m^2_{23}}{2m_{23}},\nonumber\\
|{\bm k}_1|&=&\frac{\lambda^{1/2}(m^2_{B^0_s},m^2_\phi,m^2_{23})}{2m_{23}},\nonumber\\
E_2&=&\frac{m^2_{23}+m^2_\phi-m^2_{J/\psi}}{2m_{23}},\nonumber\\
|{\bm k}_2|&=&\frac{\lambda^{1/2}(m^2_{23},m^2_\phi,m^2_{J/\psi})}{2m_{23}}
\end{eqnarray}being the energy and momentum of the $\phi$ mesons in the rest reference frame of the $J/\psi\phi$ system; $x^\ast$ is the cosine of the angle between the momenta of the $\phi$ mesons in this frame. The direct numerical evaluation shows that the integrations of Eqs.~(\ref{s12}) and (\ref{s23}) over the invariant mass intervals Eqs.~(\ref{m12}) and (\ref{m13}), respectively, give coincident  results.

\section{Application}
\label{sec5}

~

Let us apply the theoretical spectra obtained in the previous section to the description of available data \cite{LHCb}.  As it is pointed out in the Introduction, the presentation of results in terms of masses and  coupling constants of resonances with different channels is more informative than that in terms of masses and partial widths. In principle, the model includes 14 free parameters which are $m_{R_1}$, $g_{B^0_sJ/\psi R_1}$, $g_{R_1\phi\phi}$, $m_{R_2}$, $g_{B^0_sJ/\psi R_2}$, $g_{R_2\phi\phi}$, $a^{(R)}_{12}$,
$m_{X_1}$, $g_{B^0_s\phi X_1}$, $g_{X_1J/\psi\phi}$, $m_{X_2}$, $g_{B^0_s\phi X_2}$,  $g_{X_1J/\psi\phi}$, $a^{(X)}_{12}$, where $a^{(R,X)}_{12}={\rm Re}\Pi^{(R,X)}_{12}$ are taken to be constant, as explained earlier in this paper. However, the experimental $\phi\phi$ and $J/\psi\phi$ mass spectra are not normalized   so that the magnitudes of $g_{B^0_sJ/\psi R_1}$, $g_{B^0_sJ/\psi R_2}$, $g_{B^0_s\phi X_1}$, and $g_{B^0_s\phi X_2}$ have no absolute values. Hence one can obtain only the ratios of all except one to, say, $g_{B^0_sJ/\psi R_1}$. Here we fix the normalization of the LHCb data in such a way that for each spectrum, $\phi\phi$ or $J/\psi\phi$, the  plotted is the quantity
\begin{equation}\label{specexp}
f_{\rm exptl}(m)=\frac{n_{\rm bin}(m)}{\sum_{\rm bins}n_{\rm bin}(m)\Delta m},
\end{equation}
where $n_{\rm bin}(m)\Delta m$ is proportional to the number of events in the bin. Correspondingly, we plot the quantities
\begin{equation}\label{specfifi}
f_{\rm theor}(m_{\phi\phi})=\Gamma^{-1}_{\rm tot}\frac{d\Gamma}{dm_{\phi\phi}}\end{equation} and
\begin{equation}\label{specpsifi}
f_{\rm theor}(m_{J/\psi})=\Gamma^{-1}_{\rm tot}\frac{d\Gamma}{dm_{J/\psi\phi}},\end{equation} where
\begin{eqnarray}\label{Gatot}
\Gamma_{\rm
tot}&=&\int_{2m_\phi}^{m_{B^0_s}-m_{J/\psi}}\frac{d\Gamma}{dm_{\phi\phi}}dm_{\phi\phi}=\nonumber\\&&
\int_{m_\phi+m_{J/\psi}}^{m_{B^0_s}-m_{\phi}}\frac{d\Gamma}
{dm_{J/\psi\phi}}dm_{J/\psi\phi},\end{eqnarray}against the
renormalized LHCb data. Notice that $\Gamma_{\rm tot}$ is
proportional to the $B^0_s\to J/\psi\phi\phi$ decay width.  As
compared to the notations adopted in Eqs.~(\ref{s12}) and
(\ref{s23}), here  and in the figures one has $m_{\phi\phi}\equiv
m_{12}$ and $m_{J/\psi\phi}\equiv m_{23}$.

\begin{figure}
\includegraphics[width=7cm]{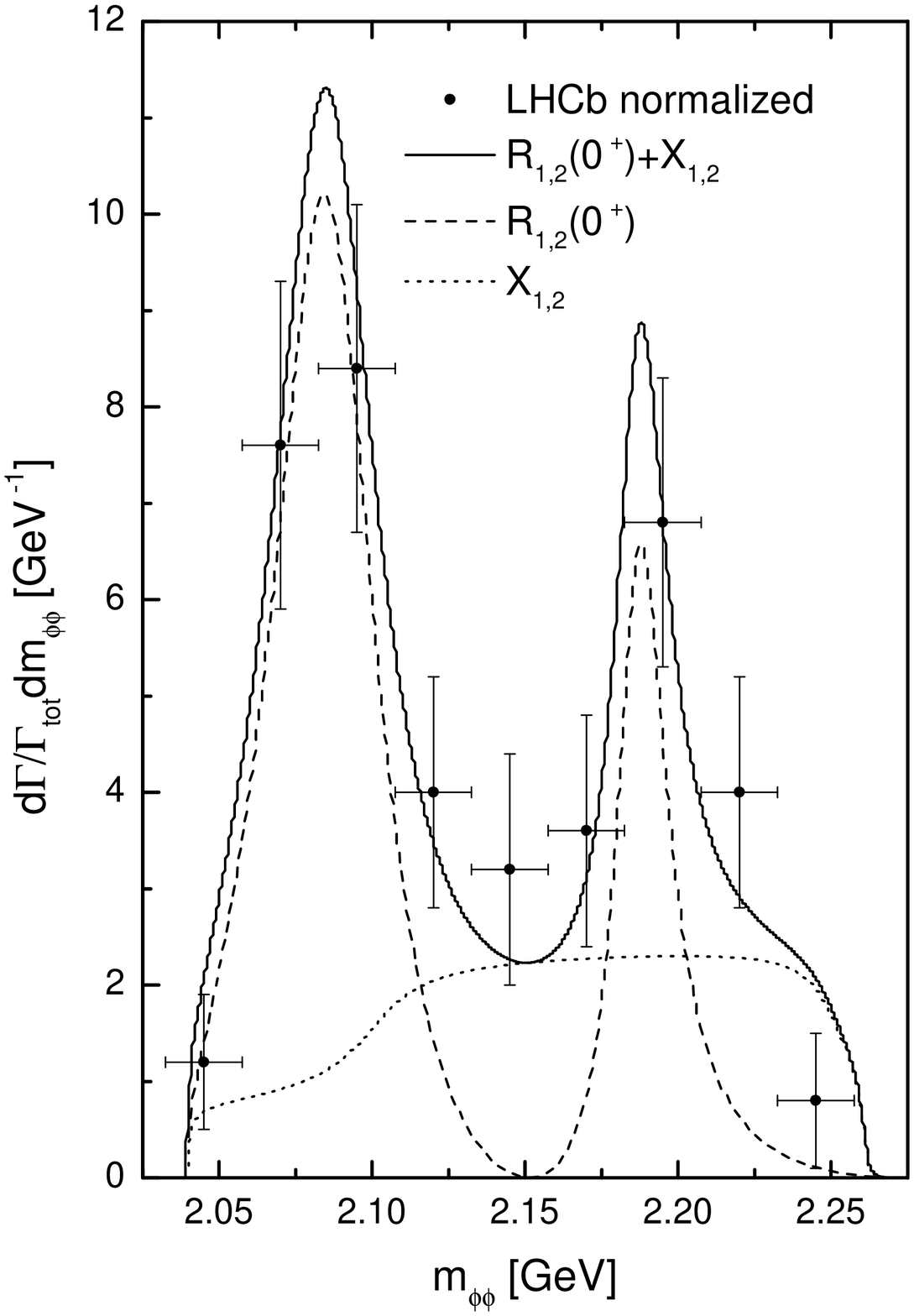}
\caption{\label{fig2}The $\phi\phi$ mass spectrum in the decay
$B^0_s\to J/\psi\phi\phi$ in the model with the $R_{1,2}$ resonance quantum numbers $J^P=0^+$ (solid curve). The contributions of the $R_{1,2}$ and $X_{1,2}$ resonances are shown with dashed and dotted lines, respectively.  LHCb data \cite{LHCb} are normalized to the unity in accord with Eq.~(\ref{specexp}).}\end{figure}

\begin{figure}
\includegraphics[width=7cm]{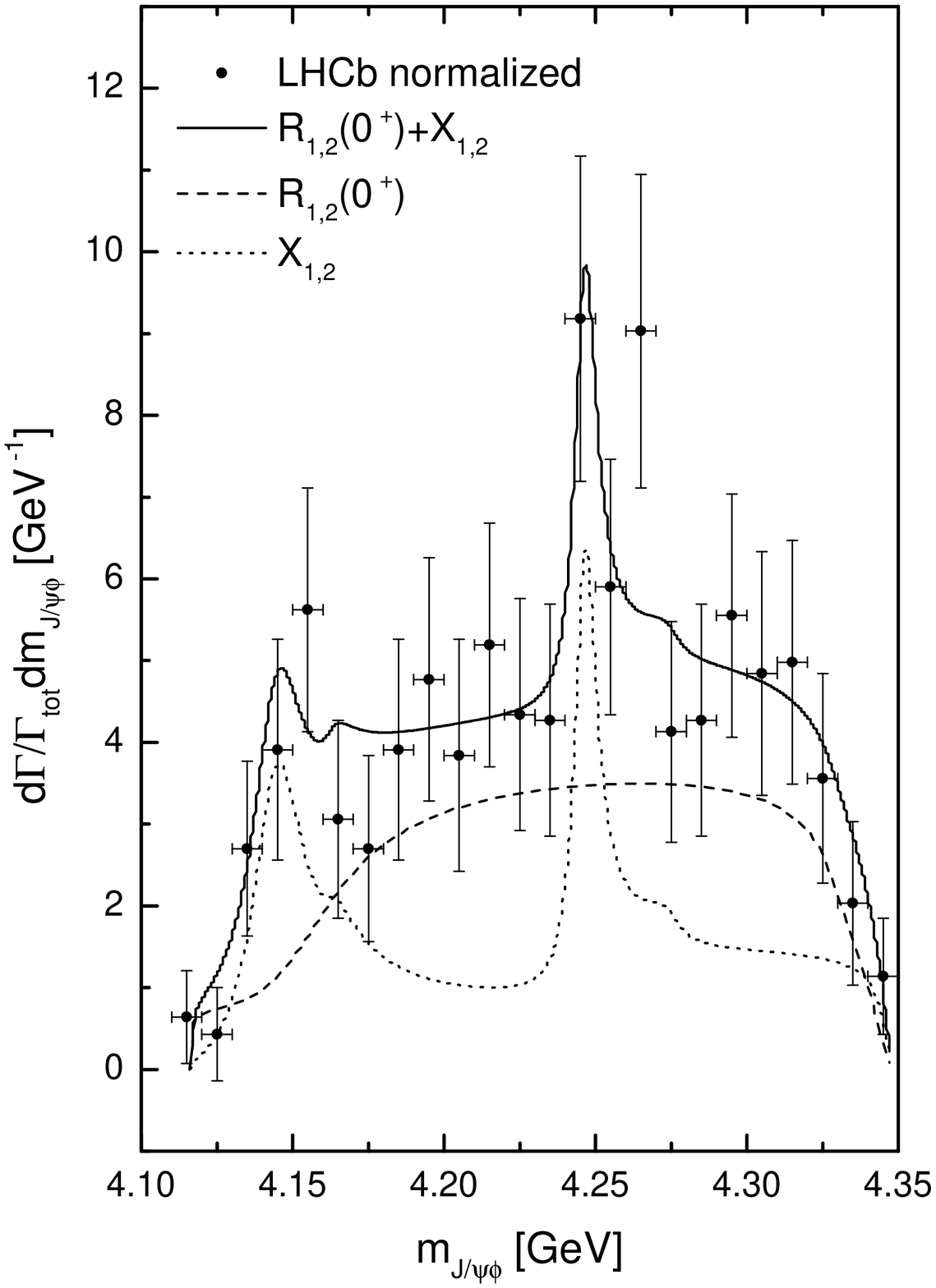}
\caption{\label{fig3}The $J/\psi\phi$ mass spectrum in the decay
$B^0_s\to J/\psi\phi\phi$ obtained in the model with the $R_{1,2}$ resonance quantum numbers $J^P=0^+$. The designations of curves are the same as in Fig.~\ref{fig2}.}\end{figure}

\begin{figure}
\includegraphics[width=7cm]{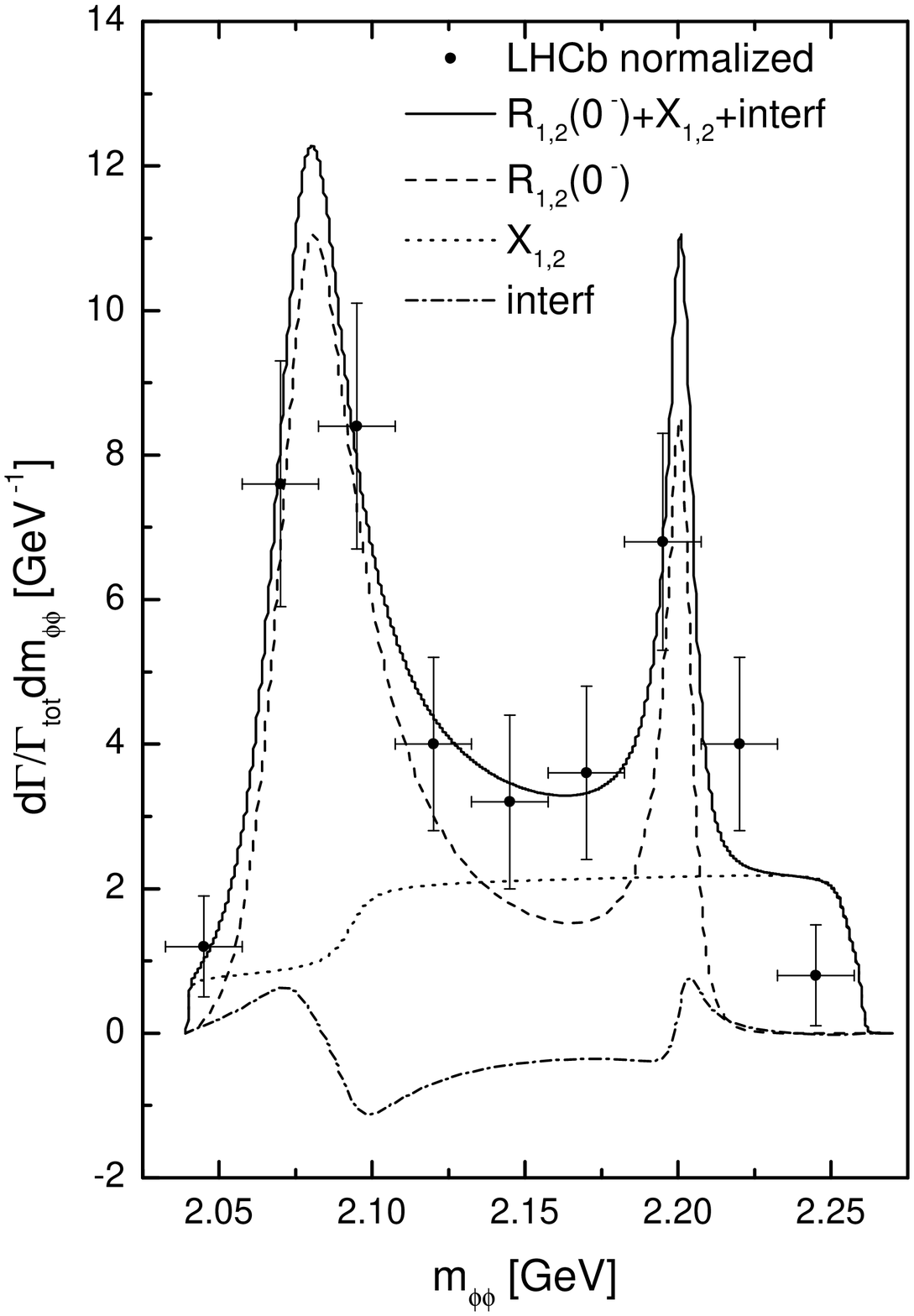}
\caption{\label{fig4}The same as in Fig.~\ref{fig2}, but in the model with the $R_{1,2}$ resonance quantum numbers $J^P=0^-$. Also shown (dot-dashed line) is the contribution of the interference term Eq.~(\ref{MB2}).}\end{figure}

\begin{figure}
\includegraphics[width=7cm]{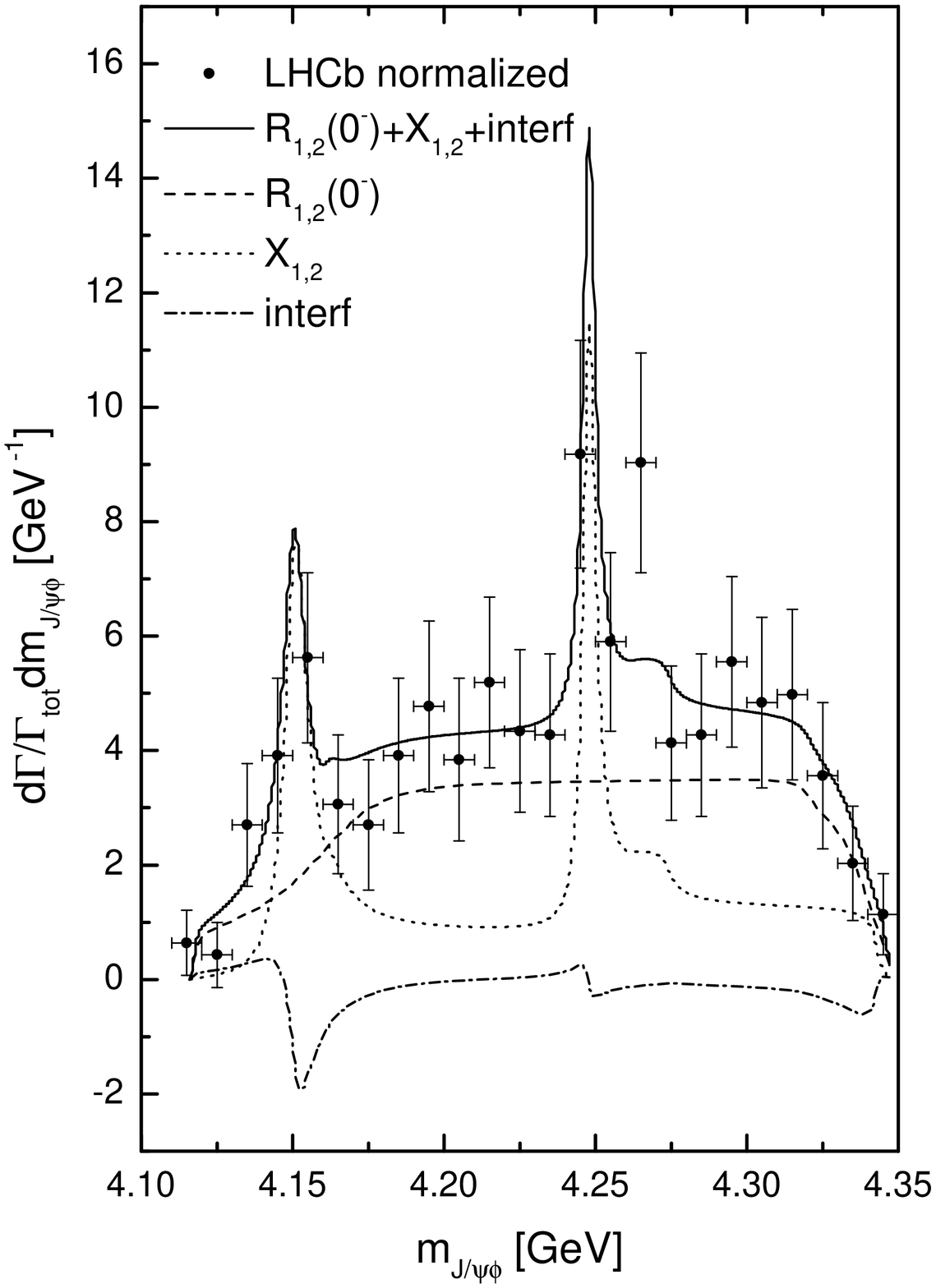}
\caption{\label{fig5}The same as in Fig.~\ref{fig3} but in the model with the $R_{1,2}$ resonance quantum numbers $J^P=0^-$.}\end{figure}

\begin{figure}
\includegraphics[width=7cm]{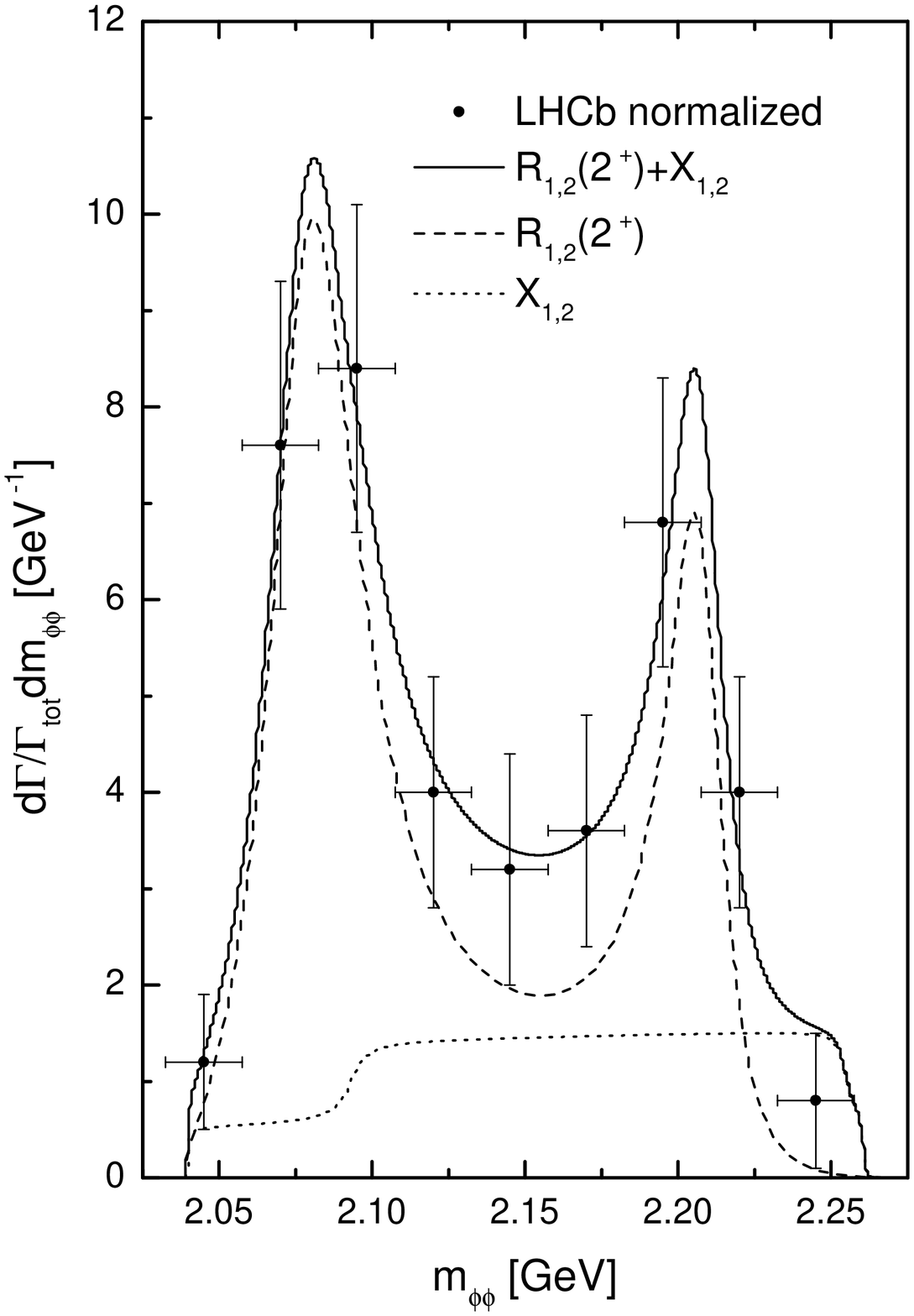}
\caption{\label{fig6}}The same as in Fig.~\ref{fig2}, but in the model with the $R_{1,2}$ resonance quantum numbers $J^P=2^+$.\end{figure}

\begin{figure}
\includegraphics[width=7cm]{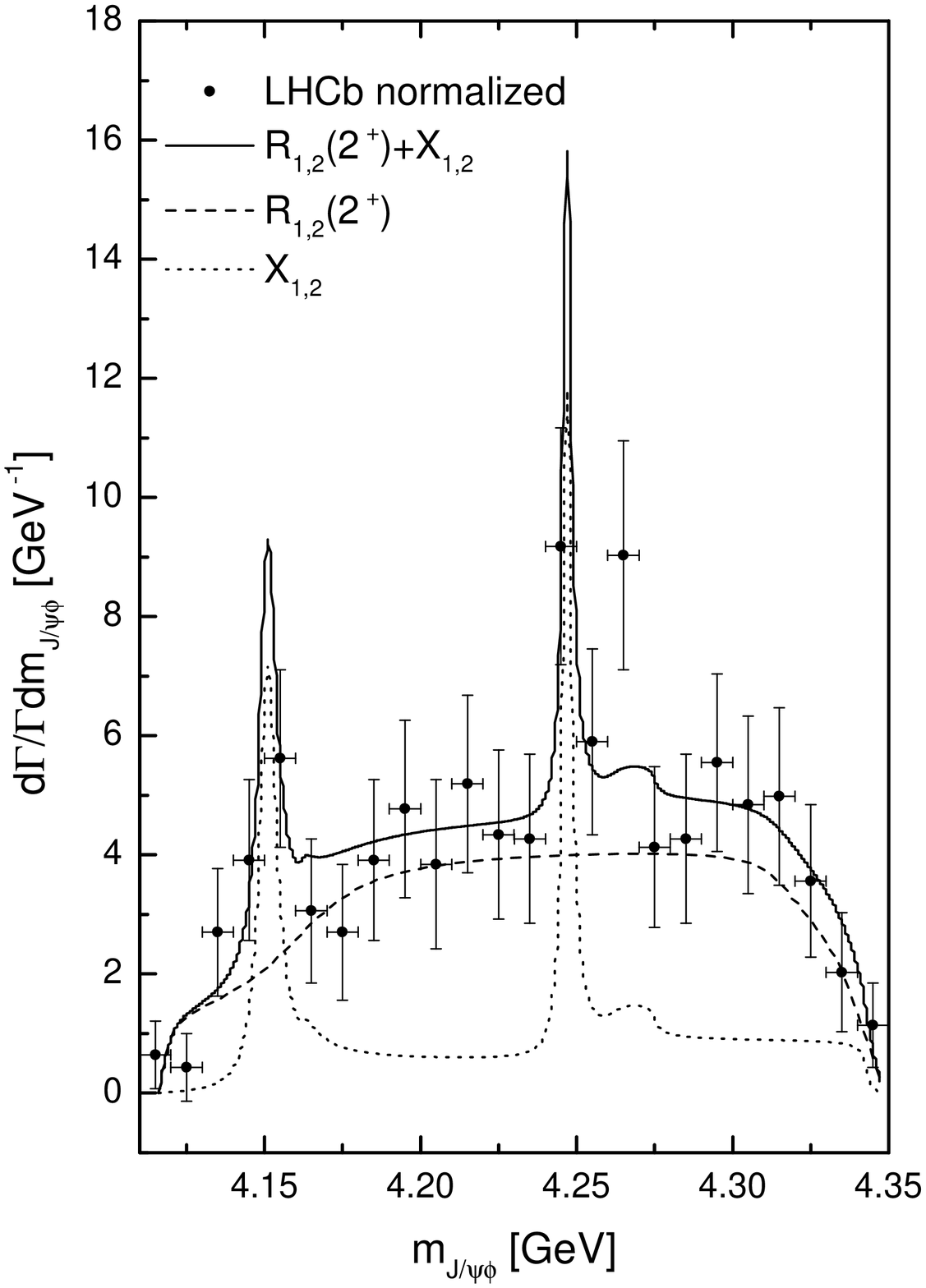}
\caption{\label{fig7}The same as in Fig.~\ref{fig3} but in the model with the $R_{1,2}$ resonance quantum numbers $J^P=2^+$.}\end{figure}

\begin{figure}
\includegraphics[width=7cm]{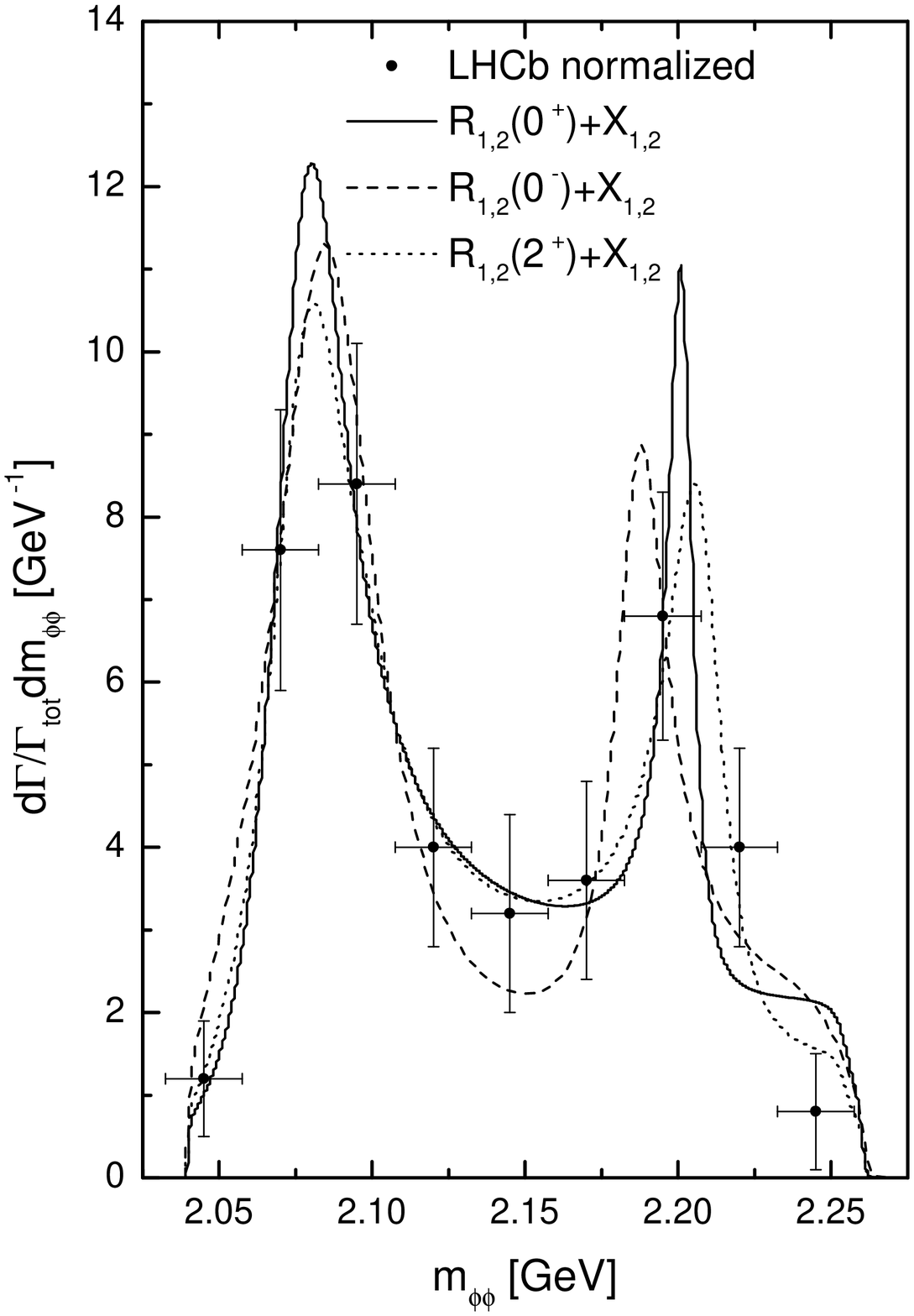}
\caption{\label{fig8}The comparison of descriptions of the $\phi\phi$ mass spectrum in the models of the $R_{1,2}$ resonances with $J^P=0^+$ (solid line), $0^-$ (dashed lone), and $2^+$ (dotted line).}\end{figure}

\begin{figure}
\includegraphics[width=7cm]{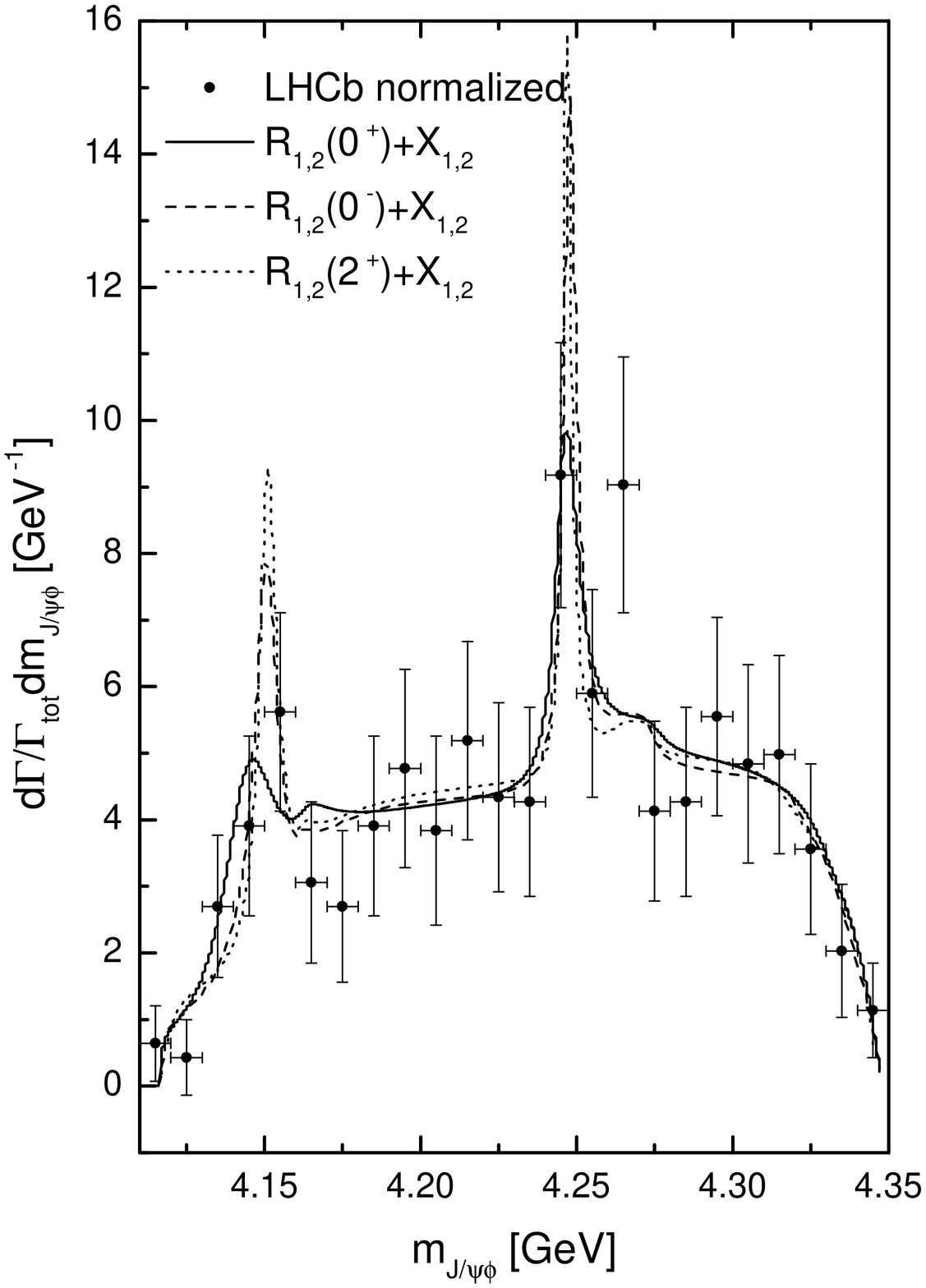}
\caption{\label{fig9}The same as in Fig.~\ref{fig8} but for the $J/\psi\phi$ mass spectrum.}\end{figure}

The results of fitting the normalized data \cite{LHCb} are
represented in Table \ref{tab1}.
\begin{table*}
\caption{\label{tab1}The resonance parameters found from fitting
the data on the $\phi\phi$ and $J/\psi\phi$ mass spectra of the decay $B^0_s\to J/\psi\phi\phi$ \cite{LHCb}.}
\begin{ruledtabular}
\begin{tabular}{llll}
 Parameter/model&$R_{1,2}=0^+$&$R_{1,2}=0^-$&$R_{1,2}=2^+$\\ \hline
 $m_{R_1}$ [GeV]&$2.089\pm0.004$&$2.088\pm0.003$&$2.081\pm0.001$\\
 $g_{R_1\phi\phi}$&$3.6\pm0.4$ GeV&$-10.5\pm0.6$ GeV$^{-1}$&$-6.1\pm0.9$ GeV\\
 $m_{R_2}$ [GeV]&$2.191\pm0.006$&$2.209\pm0.003$&$2.211\pm0.001$\\
 $\frac{g_{B^0_sJ/\psi R_2}}{g_{B^0_sJ/\psi R_1}}$&$1.1\pm0.2$&$0.6\pm0.2$&$1.5\pm0.2$\\
 $g_{R_2\phi\phi}$&$2.1\pm0.5$ GeV&$3.3\pm0.5$ GeV$^{-1}$&$3.7\pm0.2$ GeV\\
 $m_{X_1}$ [GeV]&$4.146\pm0.004$&$4.151\pm0.002$&$4.151\pm0.001$\\
 $\frac{g_{B^0_s\phi X_1}}{g_{B^0_sJ/\psi R_1}}$ [GeV]&$0.9\pm0.2$&$0.8\pm0.1$&$1.5\pm0.3$\\
 $g_{X_1J/\psi\phi}$&$-1.1\pm0.2$&$-0.7\pm0.2$&$-0.6\pm0.4$\\
 $m_{X_2}$ [GeV]&$4.247\pm0.002$&$4.248\pm0.001$&$4.247\pm0.001$\\
 $\frac{g_{B^0_s\phi X_2}}{g_{B^0_sJ/\psi R_1}}$ [GeV]&$0.9\pm0.2$&$0.9\pm0.2$&$1.6\pm0.3$\\
 $g_{X_2J/\psi\phi}$&$0.20\pm0.10$&$0.37\pm0.15$&$0.20\pm0.11$\\
 $\chi^2/n_{\rm dof}$&19/21&16/21&14/21\\
 \end{tabular}
\end{ruledtabular}
\end{table*}
When fitting, we first take into account the real parts of the polarization operators of the mixing $a^{(R)}_{12}\equiv{\rm Re}\Pi^{(R)}_{12}$ and $a^{(X)}_{12}\equiv{\rm Re}\Pi^{(X)}_{12}$ as free parameters. However, the fit chooses zero values of them, and their inclusion does not result in the lowering of $\chi^2$, hence  they are set to zero.
The corresponding curves are shown in Figs.~\ref{fig2} and  \ref{fig3} in the case of the scalar resonances $R_{1,2}=0^+$, in Figs.~\ref{fig4} and \ref{fig5} in the case of the pseudoscalar resonances $R_{1,2}=0^-$, and in Figs.~\ref{fig6} and \ref{fig7} in the case of the tensor resonances $R_{1,2}=2^+$.  Figures \ref{fig8} and \ref{fig9} demonstrate the comparison  of the curves obtained in the framework of the above three models with the LHCb data \cite{LHCb}.
In all these cases, the $X$ resonance in the $J/\psi\phi$ mass  spectrum is considered to have the quantum numbers $J^P=1^+$ \cite{LHCb1}.

When fitting experimental data,  the so-called background contribution is sometimes included. Its form is arbitrary. In the present work we have attempted to include such background by adding the pointlike amplitude with the lowest power of momenta. It looks like
\begin{eqnarray}\label{Mpoint}
M_{\rm point}&=&c_1\varepsilon_{\mu\nu\lambda\sigma}\epsilon^{J/\psi}_\mu\epsilon_{1\mu}\epsilon_{2\lambda}(k_1-k_2)_\sigma+\nonumber\\
&&c_2[(\epsilon^{J/\psi}\cdot\epsilon_1)(\epsilon_2q)+(\epsilon^{J/\psi}\cdot\epsilon_2)(\epsilon_1\cdot q)]+\nonumber\\
&&c_3[(\epsilon^{J/\psi}\cdot\epsilon_1)(\epsilon_2\cdot k_1)+(\epsilon^{J/\psi}\cdot\epsilon_2)(\epsilon_1\cdot k_2)]+\nonumber\\
&&c_4(\epsilon^{J/\psi}\cdot k)(\epsilon_1\cdot\epsilon_2),
\end{eqnarray}
where the term with $c_1$ is even under the space reflection while
those with $c_{2,3,4}$ are odd. We try to include this
contribution (upon neglecting the term $\propto c_1$ because it
contains additional power of momentum as compared with the terms
$\propto c_{2,3,4}$). However, the obtained fitting appears
unsatisfactory. The reason for this, in the present case, seems to
be the fact that, as one may observe in Figs.~\ref{fig2} -- 
\ref{fig7},
the $R_{1,2}\to\phi\phi$ decay process in Fig.~\ref{fig1} serves
as the background for the $J/\psi\phi$ mass spectrum very much
like the $X_{1,2}\to J/\psi\phi$ decay plays the same role in the
$\phi\phi$ mass spectrum.

\section{Discussion}
\label{sec6}
~

Using the coupling constants found in fits one can calculate the
central values of the  partial decay widths of the resonances
$R_{1,2}\to\phi\phi$ and $X_{1,2}\to J/\psi\phi$. They are the
following.

(A) $R_{1,2}=0^+$.
\begin{eqnarray}\label{widA}
\Gamma_{R_1\to\phi\phi}&=&40\mbox{ MeV},\nonumber\\
\Gamma_{R_2\to\phi\phi}&=&22\mbox{ MeV},\nonumber\\
\Gamma_{X_1\to J/\psi\phi}&=&21\mbox{ MeV},\nonumber\\
\Gamma_{X_2\to J/\psi\phi}&=&9\mbox{ MeV}.\nonumber\\
\end{eqnarray}
(B) $R_{1,2}=0^-$.
\begin{eqnarray}\label{widB}
\Gamma_{R_1\to\phi\phi}&=&50\mbox{ MeV},\nonumber\\
\Gamma_{R_2\to\phi\phi}&=&33\mbox{ MeV},\nonumber\\
\Gamma_{X_1\to J/\psi\phi}&=&9\mbox{ MeV},\nonumber\\
\Gamma_{X_2\to J/\psi\phi}&=&6\mbox{ MeV}.\nonumber\\
\end{eqnarray}
(C) $R_{1,2}=2^+$.
\begin{eqnarray}\label{widC}
\Gamma_{R_1\to\phi\phi}&=&36\mbox{ MeV},\nonumber\\
\Gamma_{R_2\to\phi\phi}&=&13\mbox{ MeV};\nonumber\\
\Gamma_{X_1\to J/\psi\phi}&=&7\mbox{ MeV},\nonumber\\
\Gamma_{X_2\to J/\psi\phi}&=&3\mbox{ MeV}.\nonumber\\
\end{eqnarray}
The accuracy of the $R_{1,2}$ width evaluation is about 50
percent. As for the $X_{1,2}$ resonances, their evaluated widths
spread from 7 to 21 MeV in the case of $X_1$ and from 3 to 9 MeV
in the case of $X_2$. Such wide intervals are obtained upon
evaluation with the parameters extracted from the fits with
different assumptions about spin parity of the $R_{1,2}$
resonances. In some sense such large spread  in $\Gamma_{X_{1,2}}$
can be interpreted as the model uncertainty so that the $X_{1,2}$
width evaluation is valid up to the factor of 3.

One can observe in Figs.~\ref{fig4} and \ref{fig5} that in the
variant of $R_{1,2}$ with $J^P=0^-$ the contribution of the $R-X$
interference term is relatively small. This is natural due to the
different quantum numbers of the $R_{1,2}$ and $X_{1,2}$
resonances: in the limit of their vanishing widths they do not
interfere at all.

A few words about spectroscopic identification of the $R_{1,2}$
and $X_{1,2}$ resonances considered in the present work. The
masses of the $R_1$ and $R_2$ resonances obtained from the fits
fall close to the masses of the $\eta(2100)$ and $\eta(2225)$
resonances observed by the BESIII collaboration \cite{BESIII} in the
decay $J/\psi\to\gamma\phi\phi$ but the central values of the
calculated widths are lower than those given in
Ref.~\cite{BESIII}. We attribute this to the oversimplified
assumption of the single $\phi\phi$ decay mode of the $R_{1,2}$
resonance made in the course of the present work. Also, one should
have in mind a rather large uncertainty of the widths in
Ref.~\cite{BESIII}. The same refers to the cases of $0^+$ and
$2^+$ with the possible identification $R_1\equiv f_0(2100)$ (but
without $R_2$), and $R_1\equiv f_2(2010)$ and $R_2\equiv
f_2(2300)$, respectively.

As for the $X_1$ and $X_2$ resonances in the $J/\psi\phi$ channel,
their masses obtained here from the fits fall close to the masses
of the $X(4140)$ and $X(4274)$ resonances cited in
Ref.~\cite{LHCb1}. The central values of the evaluated widths are
also lower than those given in \cite{LHCb1}. However, taking into
account the large model uncertainty up to the factor of 3 of
the evaluated widths, it seems that their values are not in
contradiction with the results of Ref.~\cite{LHCb1}.

The current theoretical interpretations of the $X(4140)$ and
$X(4274)$ resonances  as the exotic states \cite{aa16,zhu16,mai16}
rely mainly on the masses and spin-parity assignments. Further
information on their nature could be obtained from the model
predictions for the coupling constants of the considered
resonances to the pertinent final states to be compared, in turn,
with their magnitudes obtained from the data fits presented here.

\section{Conclusion}\label{sec7}
~

In the present work, the attempt is made to describe the LHCb data
\cite{LHCb} on the $\phi\phi$ and $J/\psi\phi$ mass spectra of the
decay  $B^0_s\to J/\psi\phi\phi$ in the resonance model which
takes into account the $R_{1,2}$ resonances with $J^P=0^+,0^-,$ or
$2^+$ in the $\phi\phi$ state and ones $X_{1,2}$, with $J^P=1^+$,
in the $J/\psi\phi$ state, irrespective of their nature. Taken
into account are the energy dependence of the partial widths and
the mixing inside each sector arising due to the common decay
modes. Note that the popular parametrizations of the amplitudes
neglect this kind of mixing. However, we believe  that  the mixing
due to the common decay channels being the manifestation of the
loop contribution should be taken into account because this effect
is dictated by unitarity and should exist in any effective theory.
The near-threshold kinematics has permitted one to restrict the
large number of independent Lorenz structures by a few, with the
lowest powers of momenta. The data  are still not precise enough
to make firm statements about the spectroscopy of the resonances
$R_{1,2}\to\phi\phi$ and $X_{1,2}\to J/\psi\phi$. For example, as
compared with the fits shown in Fig.~\ref{fig3}, \ref{fig5},
\ref{fig7}, and \ref{fig9}, in which the peak in the $J/\psi\phi$
mass spectrum is located at $m_{J/\psi\phi}\approx4.25$ GeV, there
are fits with the peak located at  $m_{J/\psi\phi}\approx4.26$
GeV, with equally good $\chi^2$ values. The lower experimental
point located between the higher points at the above masses does
not exclude the case of two narrow resonances while it does not
permit one to attribute the higher points to the different
resonances or to the different shoulders of a single wider
resonance. Nevertheless, the masses of the $R_1$ and $R_2$
resonances in the $\phi\phi$ mass spectrum found from the fits are
close to the values cited in the literature \cite{BESIII,pdg}. The
same refers to the resonances $X_1$ and $X_2$ in the $J/\psi\phi$
mass spectrum whose extracted masses are close to the masses of
the $X(4140)$ and $X(4274)$ resonances reported in
Refs.~\cite{LHCb1,LHCb2}. More precise data on the decay $B^0_s\to
J/\psi\phi\phi$ when (and if) appeared, together with the data on
the decay $B^+\to J/\psi\phi K^+$ \cite{LHCb1,LHCb2}, could
resolve the issue. The comparison of descriptions in the models
with different spin parity of the $R_{1,2}$ resonances presented
in Figs.~\ref{fig8} and \ref{fig9} shows no essential difference,
however, the $\chi^2$ value is lower in the case of $J^P=2^+$.
Having in mind the restricted statistics of the LHCb data
\cite{LHCb}, the inclusion of the sum of the contributions of the
$R_{1,2}$ resonances with all possible spin-parity assignments
$J^P=0^+,0^-,2^+$ seems to be premature.

\appendix
\section{Effective low momentum $X\to J/\psi\phi$ vertex}\label{A}

Let us justify the expression (\ref{Xvert}) for effective low
momentum vertex $X(1^+)\to J/\psi(1^-)\phi(1^-)$. There are three
possibilities to get $J^P=1^+$ for the $X$ resonance,
$(S,L)=(1,0),(1,2),$ and $(2,2)$, from the final state quantum
numbers of spin $S$ and angular momentum $L$. Hence, there should
be three independent Lorenz structures in the effective
Lagrangian:
\begin{eqnarray}\label{LX}
{\cal L}_{\rm
eff}&=&\frac{1}{2}\varepsilon_{\mu\nu\lambda\sigma}\left[g_1F^{(J/\psi)}_{\mu\nu}F^{(\phi)}_{\lambda\alpha}F^{(X)}_{\alpha\sigma}
+\right.\nonumber\\&&\left.g_2F^{(\phi)}_{\mu\nu}F^{(X)}_{\lambda\alpha}F^{(J/\psi)}_{\alpha\sigma}+\right.\nonumber\\&&
\left.g_3F^{(X)}_{\mu\nu}F^{(J/\psi)}_{\lambda\alpha}F^{(\phi)}_{\alpha\sigma}
\right],
\end{eqnarray}
where $F^{(A)}_{\mu\nu}=\partial_\mu V^{(A)}_\nu-\partial_\nu V^{(A)}_\mu$ stands for the field strength of the vector field $V^{(A)}_\mu$ corresponding to the vector meson $A$. Doing this in the usual way, one can obtain effective vertex from the Lagrangian (\ref{LX}). Neglecting the $D$ waves, one finds in the $X$ rest frame that
\begin{eqnarray}\label{MX1}
M_{X\to
J/\psi\phi}&=&\frac{m_X}{4}\left[\frac{m^4_X-(m^2_{J/\psi}-m^2_\phi)^2}{m^2_X}(g_1+g_2)-\right.\nonumber\\&
&\left.g_3(m^2_X-m^2_{J/\psi}-m^2_\phi)\right]\times\nonumber\\&&
({\bm\xi}^{(X)}\cdot[{\bm\xi}^{(J/\psi)}\times{\bm\xi}^{(\phi)}]),
\end{eqnarray}
where $m_A$ and ${\bm\xi}^{(A)}$ are the mass and the rest frame polarization three-vector of the meson $A$, respectively. One can denote the factors in front of the polarization structure as $g_{XJ/\psi\phi}$ and use the expression
\begin{equation}\label{MX1a}
M_{X\to
J/\psi\phi}=g_{XJ/\psi\phi}\varepsilon_{\mu\nu\lambda\sigma}p_\mu\epsilon^{(X)}_\nu\epsilon^{J/\psi}_\lambda\epsilon^{(\phi)}_\sigma
\end{equation} as the effective vertex of the $X\to J/\psi\phi$ decay in the low momentum approximation adopted throughout the paper.

\section{Mixing of resonances}\label{B}

Let us make some remarks  concerning the above expressions. As for Eqs.~(\ref{GR12}) and (\ref{GX12}), one can include more than two mixed resonances and more than one decay channel \cite{ach97}. This can be done by generalizing either Eq.~(\ref{GR12}) or  (\ref{GX12}) to the  expression
\begin{widetext}
\begin{eqnarray}\label{Ggen}
G&=&\left(
    \begin{array}{cccc}
      g_{b1}, & g_{b2}, & g_{b3}, & \cdots \\
    \end{array}
  \right)
\left(
    \begin{array}{cccc}
      D_1 & -\Pi_{12} & -\Pi_{13} & \cdots \\
      -\Pi_{12} & D_2 & \Pi_{23} & \cdots \\
      -\Pi_{13} & -\Pi_{23} & D_3 & \cdots \\
      \cdots & \cdots & \cdots & \cdots \\
    \end{array}
  \right)^{-1}\left(
                \begin{array}{c}
                  g_{1a} \\
                  g_{2a} \\
                  g_{3a} \\
                  \cdots \\
                \end{array}
              \right)
    \end{eqnarray}
    \end{widetext}
for the $a\to b$ transition amplitude through the mixed resonances $1,2,3,\cdots,N$ where $N$ is the number of mixed resonances, each  with the inverse propagator of the specific resonance like Eq.~(\ref{DRi}) or (\ref{DXi}), and with polarization operator of mixing $\Pi_{ij}\equiv\Pi_{ij}(m^2)={\rm Re}_{ij}+im\sum_ag_{ai}g_{aj}W_{i\to a}(m^2)$, where $W_{i\to a}(m^2)$ enters the partial decay width of the resonance $i$ in the form $\Gamma_{i\to a}(m^2)=g^2_{ai}W_{i\to a}(m^2)$.

The particular case is one with two  resonances mixed via the single decay mode whose partial width is $\Gamma_{1,2\to b}(m^2)=g^2_{b1,2}W(m^2)$. One can find that Eq.~(\ref{Ggen}) reduces to
\begin{widetext}
\begin{equation}\label{Gpart}
G(m^2)=\frac{(m^2_1-m^2)g_{2a}g_{b2}+(m^2_2-m^2)g_{1a}g_{b1}+(g_{1a}g_{f2}+g_{2a}g_{f1}){\rm Re}\Pi_{12}}
{(m^2_1-m^2)(m^2_2-m^2)-({\rm Re}\Pi_{12})^2-imW(m^2)\left[(m^2_1-m^2)g^2_{b2}+(m^2_2-m^2)g^2_{b1}+2g_{b1}g_{b2}{\rm Re}\Pi_{12})\right]}.
\end{equation}
\end{widetext}
In fact, it is this form that is used in the present work. One may observe  that if Re$\Pi_{12}\to0$ then in the vicinity of  $m=m_1$ ($m=m_2$) Eq.~(\ref{Gpart}) reduces to the simple Breit-Wigner form Eq.~(\ref{BW}) for the resonance 1 (2).

\end{document}